\documentclass[11pt,a4paper]{article}
\usepackage{jcappub,float,amsmath,amssymb,amsfonts,graphicx,threeparttable,hyperref,url,epstopdf}
\usepackage{enumerate,multirow}

\usepackage[font=small,labelfont=bf]{caption}

\title{Joint Planck and WMAP Assessment of Low CMB Multipoles}
\author[a]{Asif Iqbal,}
\author[b]{Jayanti Prasad,}
\author[b]{Tarun Souradeep} 
\author[a,b]{and Manzoor A. Malik}
\affiliation[a]{Department of physics, University of Kashmir, Hazratbal, \\
Srinagar, Jammu and Kashmir 190006, India}
\affiliation[b]{Inter-University Center for Astronomy and Astrophysics, Post Bag 4, Ganeshkhind, \\
Pune 411007, India}

\emailAdd{asif.iqbal31@gmail.com}
\emailAdd{jayanti@iucaa.ernet.in}
\emailAdd{tarun@iucaa.ernet.in}
\emailAdd{mmalik@kashmiruniversity.ac.in}    
\abstract{The remarkable progress in cosmic microwave background (CMB) studies 
over past decade has led to the era of precision cosmology in striking agreement with the $\Lambda$CDM model. 
However, the lack of power in the CMB temperature anisotropies at large angular scales (low-$\ell$), 
as has been confirmed by the recent Planck data also (up to $\ell=40$), 
although  statistically not very strong (less than $3\sigma$),  is  still an open problem. 
One can avoid to seek an explanation for this problem by  attributing the lack of power to cosmic variance or
can look for explanations i.e., different inflationary potentials or  
initial conditions for inflation to begin with, non-trivial topology, ISW effect etc. 
Features in the primordial power spectrum (PPS) motivated by the early universe physics  
has been the most common  solution  to address this  problem. In the present work we also follow this approach and  
consider a set of PPS  which have features and constrain 
the  parameters of those using WMAP 9 year and Planck data employing Markov-Chain Monte Carlo (MCMC) analysis.
The prominent feature of all the  models of PPS that we consider is an infra-red cut off which 
leads to suppression of power at  large angular scales.  We consider models of PPS with maximum three extra parameters and use Akaike information criterion ($AIC$) and Bayesian information criterion ($BIC$)  of model 
selection to compare the models. For most models,  we find good constraints for the cut off scale $k_c$, however, for other parameters our constraints are not that good. 
We find that sharp cut off model gives best likelihood value for the WMAP 9 year data, but is as good as power law model according to $AIC$. 
For the joint WMAP 9 + Planck data set, Starobinsky  model is slightly preferred by $AIC$ which is also able to produce CMB power suppression up to $\ell\leq30$ to some extent.
 However, using $BIC$ criteria, one finds model(s) with least number of parameters (power law model) are always preferred. }
\begin{document}
\maketitle

\section{Introduction}

Anisotropies in the temperature and polarization  of cosmic microwave background (CMB)  have information 
about the early universe and can be used to study the primordial fluctuations generated during inflation which lead to 
structure formation in the universe \cite{1981PhRvD..23..347G,1982PhLB..115..295H,1982PhLB..117..175S,1982PhRvL..49.1110G,1993PhR...231....1L,2000cils.book.....L}.
CMB anisotropies are believed to be sourced by quantum fluctuations generated during inflation 
\cite{1982PhLB..115..295H,1982PhLB..117..175S,1994PhRvD..50.7222L,2005hep.th....3203L}. 
Thus by investigation of the CMB angular power spectrum,  which completely characterizes CMB anisotropies, one is able to probe 
the physics of the very early universe.  

So far most CMB experiments indicate that CMB anisotropies are statistically isotropic and Gaussian and so  can be completely 
characterized by their two-point correlations or power spectrum \cite{2013ApJS..208...20B,2013ApJS..208...19H,2014A&A...571A..16P}.
Although, almost all the CMB observations confirm that the six parameter $\Lambda$CDM  cosmological  model  best fits the observed data, still 
there are some anomalies which have always been  present from COBE to Planck. One of such anomalies has been the lack of power in 
the CMB-TT power spectrum ($C_l^{TT}$) at large angular scales or low-$\ell$  \cite{1994PhRvL..73.1882J,2011ApJS..192...17B,2014A&A...571A..15P}.
Recently, \cite{2014JCAP...01..043H} studied the consistency of the 
standard $\Lambda$CDM model with the Planck data using the Crossing statistic \cite{2011JCAP...08..017S,2012JCAP...05..024S,2012JCAP...08..002S}.
 Their results indicate that the Planck data is consistent to the concordance $\Lambda$CDM only at 2-3$\sigma$ confidence level 
and lack of power at both high and low $\ell$'s with respect to concordance model.
The low power at large angular scales can be attributed to cosmic variance \cite{2003MNRAS.346L..26E}, still there have been efforts to  explain this anomaly 
by changing the potential of inflation field (for a comprehensive review check \cite{2013arXiv1303.3787M}), considering different initial 
conditions at the beginning of inflation \cite{2003JCAP...07..002C,2006PhRvD..74l3006B,2007PhRvD..76f3512P,2008PhRvD..77h3501W,2014arXiv1412.7093D,2014arXiv1408.5904B,2014JCAP...12..030C,Berera1998}, ISW effect \cite{2014JCAP...02..002D}, 
spatial curvature \cite{2003MNRAS.343L..95E}, non-trivial topology \cite{2003Natur.425..593L}, geometry \cite{2006PhRvL..97m1302C,2007PhRvD..76f3007C},
 violation of statistical anisotropies \cite{2003ApJ...597L...5H}, cosmological-constant type dark energy during the inflation \cite{2004PhRvD..70h3003G}, 
bounce from contracting phase to inflation \cite{2004PhRvD..69j3520P,Liu2013}, production of primordial micro black holes (MBH) remnants in the very early universe \cite{Scardigli2011}, 
hemispherical anisotropy and non-gaussianity \cite{McDonald2014b,McDonald2014}, string theory \cite{2014arXiv1411.6396K}, loop
quantum cosmology \cite{2014CQGra..31e3001B} etc.
Although, inflationary $\Lambda$CDM model with almost scale-invariant power spectrum  has emerged most successful model from the recent observations,
it is important to note it does not uniquely confirm the generic picture of the universe and the generalization 
of primordial power spectrum having additional features like cut off, oscillations would be crucial in identifying specific inflationary models.

There are two main approaches which have been followed to probe the primordial power spectrum of curvature fluctuations generated during 
inflation  from the CMB anisotropies. In the first approach no specific model of PPS is considered and the shape of PPS 
is directly reconstructed from the data by deconvolution using different techniques 
\cite{2001PhRvD..63d3009H,2003ApJ...599....1M,2003MNRAS.342L..72B,Shafieloo2004,2007PhRvD..75l3502S,2009MNRAS.400.1075B,2010JCAP...01..016N,2014A&A...571L...1L}. 
The main disadvantage of this approach is that the angular power spectrum does not have all the information about
PPS due to nature of transformation kernel and some form of regularization may be needed which penalizes models
which have features not desired. Maximum Entropy method \cite{2013PhRvD..88b3522G} has also been applied for this purpose. 
Planck team has  used a form of regularization which penalizes any model of PPS which  
deviates  from  a straight line and has non-zero power at those scales  at which there are no constrains \cite{2014A&A...571A..22P}. 
The second approach which has been used to probe the primordial power spectrum from the CMB data has been to
consider physically motivated models of the early universe, represented by a PPS with features, and estimate 
the parameters of those from CMB data 
\cite{1982PhRvD..26.1231V,1992JETPL..55..489S,2003JCAP...07..002C,2004PhRvD..69j3520P,2006PhRvD..74d3518S,2007PhRvD..76f3512P}. 
Some of the models of PPS are inspired from the change in physical conditions during inflation represented by 
change in the potential of inflation during slow roll or initial conditions at the beginning of inflation. 
In the present work, we consider a set of models of PPS which were studied in \cite{2006PhRvD..74d3518S} and try to
constrain the parameters of those from the WMAP 9 year and Planck data. We add a couple of other models 
also to our analysis. All the models we consider have a common feature that they
all have an infrared cut off in power at large scales and match perfectly with the standard power law model at small scales. 

The plan of this paper is as follows: In Section 2, we give a brief outline of inflationary framework 
to introduce primordial power spectrum and its parameters and also give a short introduction of the 
models (PPS) we consider for our analysis. In section 3, we present the results of our analysis 
in the form of the best fit model parameters of PPS. We also give a comparison of our models
using Akaike information criterion ($AIC$) and Bayesian information criterion ($BIC$)  in this section. Discussion and conclusions of our
work are given in the last section.

\section{Inflationary fluctuations}

Inflation is characterized by a phase in the early universe when the energy density of the universe is dominated by a scalar field $\phi$ 
that can be characterized by a perfect fluid with the diagonal components of the energy-momentum tensor given by the 
energy density  $\rho_{\phi}$ and pressure density $p_{\phi}$ respectively  \cite{1994PhRvD..50.7222L,2005hep.th....3203L,2013arXiv1303.3787M}. 
\begin{equation}
\rho_{\phi} = \frac{1}{2} {\dot \phi}^2 + V (\phi),
\end{equation}
and 
\begin{equation}
p_{\phi} = \frac{1}{2} {\dot \phi}^2 - V (\phi),
\end{equation} 
where $V(\phi)$ is the potential energy of the scalar field. 

The dynamics of the scalar field that leads to inflation is governed by the following equation (in FRW  case):
\begin{equation}
{\ddot \phi} + 3 H {\dot \phi} + V'(\phi) = 0 
\label{eqn1:inf}
\end{equation}
and 
\begin{equation}
H^2 = \frac{1}{3M_{Pl}^2} \left (\frac{1}{2} {\dot \phi}^2 + V (\phi)  \right).
\end{equation}

Slow-roll inflation is characterized by two parameters $ \epsilon$ and $\eta$ which are
defined as :
\begin{equation}
\epsilon   = -\frac{\dot H}{H^2}  = \frac{M_{Pl}^2}{2} \left(\frac{V'(\phi)}{V(\phi)} \right)^2
\end{equation}
and
\begin{equation}
\eta =  \epsilon  + \delta = M_{Pl}^2 \left ( \frac{V''(\phi)}{V(\phi)} \right),
\end{equation}
where 
\begin{equation}
\delta = - \frac{\ddot \phi}{H {\dot \phi}}.
\end{equation}
For inflation to happen we need $ \epsilon << 1$ and $ |\eta| << 1$. 

During inflation fluctuation $\delta \phi$  in the scalar field $\phi$ lead to fluctuation ${\cal R}$ 
in the spatial curvature which  are characterized by their  two point 
 correlation function under common assumptions (homogeneity \& isotropy): 

\begin{eqnarray}
 \langle {\cal R}^{*}({\bf k}) {\cal R}({\bf k}')  \rangle \, = \,  \delta^3({\bf k} - {\bf k}')\Delta_{\cal R}^2(k) \, , \qquad \mathcal{P}_o(k) \equiv \frac{k^3}{2\pi^2} \, \Delta_{\cal R}^2(k)\, ,
\end{eqnarray}

where the angular brackets denote an ensemble average, $\delta$ is the Dirac delta function and $\mathcal{P}_o(k)$ is called primordial scalar power spectrum. 
In the standard $\Lambda$CDM cosmology the shape of the primordial power spectrum in its simplest form can be expressed in power-law parameterization.
 This model is referred to as Power Law model and can be obtained at leading order slow-roll approximation of the single-inflation field \cite{1995PhRvD..52.1739K}:
\begin{equation}
\mathcal{P}_o(k)  =  A_S \left( \frac{k}{k_0} \right)^{n_s-1},
\label{plspec}
\end{equation}
where $n_s$ is called spectral index (tilt parameter) and is expected to be close to 1, $A_s$ is spectral amplitude, 
$k_0$ is the scalar pivot which is set equal to $0.05$ Mpc$^{-1}$ throughout this work. 
The scalar primordial power spectrum parameters can be calculated in terms of slow roll parameters $(\epsilon,\, \eta)$ as
\begin{equation}
A_s\simeq \frac{V}{24\pi^2\epsilon M_P^4}, \quad n_s\, \simeq \, 1 +2\eta-6\epsilon.
\end{equation} 

In addition, to the scalar primordial power spectrum, inflation also predicts a tensor spectrum $\mathcal{P}_t(k)$ due to 
gravity-wave (tensor) perturbations which is usually written in the form
\begin{equation}
\ln \mathcal{P}_t(k)\, = \, \ln A_t\,+ n_t\ln \left(\frac{k}{k_0}\right),
\end{equation}
where $A_t$ and $n_t$ are the tensor amplitude and spectral index respectively. In terms of slow roll parameter these can be written as
\begin{equation}
 A_t\simeq \frac{3V}{2\pi^2M_P^2}, \quad r\equiv \frac{\mathcal{P}_t}{\mathcal{P}_o}\simeq16\epsilon, \quad n_t\simeq-\frac{r}{8}.
\end{equation}
The major contribution to $P_t(k)$ comes form the B-mode polarization and in light of recent questions regarding the claims of the BICEP2 results \cite{2014PhRvL.112x1101A,2014ApJ...789L..29L,Adam2014}, we will not use B-mode BICEP2 
polarization data in our analysis and therefore assume $r = 0$ (or $P_t(k) = 0$) (i.e consider scalar perturbations only).

The inflation theory predict a temperature fluctuations  to be  statistically isotropic with very nearly Gaussian of zero mean, consistent with
 current observations. It is customary to represent theoretical and experiment temperature power spectrum in terms of spherical harmonics as \cite{Hu2002}
\begin{equation}
 \frac{\Delta T(\hat{n})}{T}=\sum_{\ell=1}^{\infty}\sum_{m=-\ell}^{m=\ell}\,a_{\ell m}\,Y_{\ell m}(\hat{n}),
\end{equation}
where $\hat{n}\equiv(\theta,\phi)$ is a unit direction vector on the sky, $a_{\ell m}$ are complex quantities and $Y_{\ell m}(\hat{n})$ are 
normalized spherical harmonics.
Assuming CMB fluctuations to be Gaussian distributed, then each $a_{lm}$ is independent with exception equal to zero and Gaussian distributed: 
\begin{equation}
<a_{lm}^*a_{l'm'}> = \delta_{ll'} \, \delta_{mm'}\,C_l,
\end{equation}
where $C_l$ is called the angular power spectrum. In practice, CMB angular power spectrum $C_{\ell}$  is computed using the  two-point angular correlation function 
\begin{equation} 
C(\hat{n}_{1}\cdot\hat{n}_{2})=\left< \frac{\Delta T}{T}(\hat{n}_{1})\frac{\Delta T}{T}(\hat{n}_{2})\right>=\sum_{\ell=2}^{\infty}\, \frac{2\ell+1}{4\pi}\,C_{\ell}\,P_{\ell}(\hat{n}_{1}\cdot\hat{n}_{2}),
 \end{equation}
where $P_{\ell}$ is the Legendre polynomials. The measured angular power spectrum $C_l$ is a robust cosmological probe in constraining cosmological models, the position and amplitude of the peaks 
being very sensitive to important cosmological parameters. Since Thomson scattering of an anisotropic radiation field also generates linear polarization \cite{Hu2002}, there are also angular power spectrum due to the polarization. 
The polarization anisotropies have a different dependence on cosmological parameters than that for temperature power spectrum and can provide a way to break degeneracies in various cosmological parameters. 
Moreover, since polarization data is free from ISW effect one can easily separate out the contribution of low CMB power due to ISW and infrared cut off.  
The polarization can be  divided into parts that come from curl (B-mode) and  divergence (E-mode) yielding four independent angular power spectra as $C^{TT}_{\ell}$,$C^{EE}_{\ell}$, $C^{TE}_{\ell}$,
$C^{BB}_{\ell}$. The initial power spectrum $ \mathcal{P}(k)$  is related to the angular power spectrum $C_{\ell}$ through
\begin{equation}
C_\ell^{{XX'}}  \propto \int {\rm d}{\rm ln} k \, \mathcal{P}(k) \, T_\ell^{X} (k) \, T_\ell^{X'} (k),
\end{equation}
where $T_\ell^X(k)$ is the transfer function  with $X$ representing the CMB temperature or polarization. 

One of the noteworthy outcomes from recent cosmological results, especially from WMAP and Planck,  is the 
possibility of obtaining structural form of the  primordial power spectrum  \cite{DK2014,2013JCAP...12..035H,2014PhRvD..90b3544H,2014JCAP...01..025H}, which in turn has potential to differentiate strongly between
various inflationary models dominating early universe physics. 
The most commonly  used primordial spectrum is almost scale-invariant power law during inflation which went on to produce the observed structure in the CMB. 
However, different inflationary models readily accommodate different primordial spectra with radical departures from this simple picture, especially at low $k$.  Moreover, recent results 
of WMAP and Planck have confirmed the general picture of the primordial power spectrum  having a  suppression at low $k$ which could not be explained by scale-invariant power law model. 
Deconvolution of CMB data strongly favors  a cut off around horizon scale $ 0.00001$Mpc$^{-1}< k_{c}< 0.0009$Mpc$^{-1}$ followed by a bump in a primordial power spectrum  \cite{2013JCAP...12..035H,2014PhRvD..90b3544H}.
Motivated by the fact that primordial power spectra with a cut off should give better likelihood than scale free power law model, we will next 
point out various primordial power spectra which have cut off at low $k$ arising due to the physics in the initial phase of inflationary models. 

In the present work we consider models of primordial power spectrum (PPS) which suppress power at large scales (small-$k$) and agree 
at small scales with the standard power law model since our aim here is to explain the deficiency of power at large angular scales in CMB-TT power spectrum.
By considering models with a large number of fitting parameters it becomes easier to fit the data, however, any 
method of model comparison must penalize models with a large number of fitting parameters. In the present work, 
we consider models with at the most three extra parameters of  PPS (apart from two usual parameters $A_s$ and $n_s$). For model selection we will 
use $AIC$ and $BIC$ in the next section.

\subsection{Model 1 : Power law  (PL)}
We consider the standard power law power spectrum ``$\mathcal{P}_o(k)$'' characterized by two parameters, spectral index ($n_s$) and 
amplitude $A_s$ at some pivot scale $k_0$.  Since this model is a  part  of the standard six parameters cosmological model
therefore we compare the improvement in the likelihood as compared to this model. ${\mathcal P}_0(k)$ is give by Eq.~(\ref{plspec}) and can be rewritten as:
\begin{equation}
\ln {\mathcal P}(k) = \ln A_s  + (n_s-1)  \ln  \left ( \frac{k}{k_0} \right).  
\end{equation}
All the models we consider in the present work can be written as modulation over the power law model:
\begin{equation}
{\mathcal P}(k) = {\mathcal P}_0(k) \times {\mathcal F}(k,{\bf \Theta}),
\end{equation}
where ${\mathcal F}(k,{\bf \Theta})$  is 
the  ``modulation''  part and ${\bf \Theta}$ is a vector which characterizes the extra parameters.

\subsection{Model 2 : Running spectral index (RN)}
Scale dependent spectral index $n_s$, as  characterized by an extra parameter $\alpha_s$  called ``running index'',
has been a part of the extension of the standard six parameter cosmological model 
and is well motivated in the inflationary framework \cite{1995PhRvD..52.1739K,2006JCAP...09..010E,2011JCAP...01..026K,2014PhLB..735..176C}. 
 In the slow-roll approximation, $\alpha_s$, is second order term and is of the  order of $10^{-3}$ and therefore was assumed to be zero for
the power law model discussed in previous section. $\alpha_s$ can be calculated in terms of slow roll parameters as
\begin{equation}
\alpha_s\simeq-2\xi +16\epsilon \eta -24\epsilon^2,
\end{equation} 
where the slow roll parameter $\xi$ is related to the third derivative of the inflationary potential  $V(\phi)$ in the following way
\begin{equation}
\xi=M_{Pl}^4\frac{V'V'''}{V^2}.
\end{equation}
Although, larger value of $\alpha_s$ could produce suppression of CMB power, but  
sizable value of $\alpha_s$  will amount to violation of slow roll approximation. However, there are certain models  \cite{Ballesteros2006,Ballesteros2008,2006JCAP...09..010E,2011JCAP...01..026K,2014PhLB..735..176C} 
where $\alpha_s$ can be large, while still respecting the slow-roll approximation. Therefore, in our analysis we also consider a model with non-zero $\alpha_s$.

Current CMB observations slightly favor a non-zero running model of PPS over the standard power law PPS and in 
the present work we try to constrain the parameter $\alpha_s$ with the WMAP 9 year and Planck data. We use the standard parameterization 
for the running which is given by the following equation.
\begin{equation}
\ln \mathcal{P}(k)\, = \,\ln A_s +\,(n_s-1)\ln \left(\frac{k}{k_0}\right) +\frac{\alpha_s}{2}\left[\ln \left(\frac{k}{k_0}\right)\right]^2.
\label{plspec2}
\end{equation}

\subsection{Model 3 : Sharp cut off  (SC)}
This model assumes that there is sharp cut off in the primordial power spectrum at large scale:

\[
    P(k)= 
\begin{cases}
     A_s \left(\frac{k}{k_c} \right)^{n_s-1},& \text{for}~~ k > k_c \\
    0,              & \text{otherwise.}
\end{cases}
\]
This model was considered in \cite{2006MNRAS.369.1123B,2009MNRAS.400.1075B} and constraints were found for the cut off scale.
One of the interesting features of this model is that it has just one extra parameter and fits the 
data as closely as the exponential cut off model with two extra parameters which is discussed in Sec.~(\ref{section2.5}).
\subsection{Model 4 : Pre-inflationary radiation domination (PIR)}
In this model we take into account the effect of a pre-inflation radiation-dominated era which can lead to modulations in the primordial power spectrum \cite{1982PhRvD..26.1231V,2007PhRvD..76f3512P,2008PhRvD..77h3501W}. 
The transition from a pre-inflation radiation-dominated  phase to de-Sitter universe was first studied by Vilenkin \& Ford \cite{1982PhRvD..26.1231V}: 
\begin{equation}
\mathcal{P}(k) = A_s\,k^{1-n_s}\,\frac{1}{4 y^4}\left| e^{-2 i y} (1+2 i y)-1 - 2 y^2\right|^2, 
\end{equation} 
where $y={k}/{k_c}$. The cut off scale $k_c$ is set by the Hubble parameter and is proportional to $k^2$. Here, current horizon crosses the horizon around 
the onset of inflation. This model produces cut off followed by a bump like feature in the primordial power spectrum.

\subsection{Model 5 : Pre-inflationary kinetic domination (PIK)}

We also consider a model given in \cite{2003JCAP...07..002C} which also produces cut off  due to possible existence of a kinetic stage in the pre-inflationary era, 
where the velocity of the scalar field was not negligibly small. In order to affect the low-$\ell$ multipoles, this stage should occur very close to the beginning of
the last 65 e-fold period of inflation. If scales corresponding to the current horizon have exited the horizon around the onset of inflation then this could cause a 
significant drop on the large angular scales of the primordial spectrum.  
 Here the inflation potential is quadratic
\begin{equation}
 V(\phi)=\frac{m^2_{\phi}\phi^2}{2},
\end{equation}
with initial conditions given by $\phi_{in}=18M_p$, $(d\phi/dt)_{in}\simeq-m_{\phi}\phi_{in}$.
The form of primordial perturbations for pre-inflationary kinetic domination model can be expressed as
\begin{equation}
\mathcal{P}(k) \, =\, {{H^2_{inf}}\over{2\pi^2}} \, k \, {\mid A -  B \mid}^2 ,
\label{eqn:pk}
\end{equation} 
where
\begin{eqnarray}
A \, = \, \frac{e^{-ik/H_{inf}}}{\sqrt{32 \, H_{inf}/\pi}} \left[ {\mathcal{H}}_0^{(2)} \left(\frac{k}{2H_{inf}} \right) \, - \, \left( \frac{H_{inf}}{k}\, + \, i \right) \,
{\mathcal{H}}_1^{(2)} \left( \frac{k}{2H_{inf}} \right) \right], \nonumber
\end{eqnarray} 
\begin{eqnarray}
B \, = \, \frac{e^{ik/H_{inf}}}{\sqrt{32 \, H_{inf}/\pi}} \left[ {\mathcal{H}}_0^{(2)} \left(
\frac{k}{2H_{inf}} \right) \, - \, \left( \frac{H_{inf}}{k}\, - \, i \right) \,
{\mathcal{H}}_1^{(2)} \left( \frac{k}{2H_{inf}} \right) \right],  \nonumber
\end{eqnarray} 
$H_{inf}$ denotes the Hubble parameter in physical units during inflation, $\mathcal{H}_0^{(2)}$ and $\mathcal{H}_1^{(2)}$ denote the Hankel function 
of the second kind with order 0 and 1, respectively. Here the cut off is proportional to $k^3$. 
However, this model has a scale invariant primordial power spectrum for large $k$ and therefore is strongly disfavored by the current data, despite producing  low CMB power.


If we consider that quantum fluctuations originate in the Bunch-Davies vacuum as is considered in 
\cite{2007PhRvD..76f3512P} also~\footnote{Note that the validity of imposing Bunch-Davies vacuum 
in Kinetic Domination regime at all scales is questionable but this assumption does provide a specific form of  model PPS to study.}, 
then the primordial power spectrum can be rewritten as:

\begin{equation}
\mathcal{P}(k) \, =\, A_s' \left( \frac{k}{k_0} \right)^{n_s-1}   {{H^2_{inf}}\over{2\pi^2}} \, k \, {\mid A -  B \mid}^2, 
\end{equation} 
with 
\begin{equation}
A_s  =  A_s'  {{H^2_{inf}}\over{2\pi^2}} \, k_0 \, {\mid A(k_0) -  B(k_0) \mid}^2.
\end{equation}
where $k_0$ is the pivot scale. 

This model also has one extra parameter $H_{inf}$ which we constrain from WMAP 9 + Planck data. 

\subsection{Model 6 : Exponential cut off  (EC)}\label{section2.5}
The primordial power spectrum  which has  lesser power at low $k$ can also be approximated by imposing an
 exponential cut off at $k < k_c$ \cite{2003JCAP...07..002C,2003JCAP...09..010C,2006MNRAS.369.1123B,2010PhRvD..82l3009G} on the power law model $\mathcal{P}_o(k)$:
\begin{equation}
\mathcal{P}(k)\,=\, \mathcal{P}_o(k)\,\left[1-e^{-(k/k_c)^{\alpha}}\right],
\end{equation}
 where $\alpha$ is a measure of the steepness of the cut off. On small angular scales, this parameterization behaves as simple power law model 
and qualitative features in the CMB power spectrum that determine the constraints on the cosmological parameters are not affected  except at the low-$\ell$ multipoles. 
\subsection{Model 7 : Starobinsky (SB)}
Another model which predicts a step-like feature in $\mathcal{P}(k)$ was proposed by Starobinsky \cite{1992JETPL..55..489S}, which assumes that there is a sharp change in slope of
potential of the scalar field $V(\phi)$ at certain $\phi_0$ which controls the inflationary stage. The general form of scalar field potential which has a rapid change  for such cases can be expressed as
\begin{eqnarray}
V(\phi)=\left\{
                \begin{array}{lr}
                   V_0+A_+ \, \left(\phi-\phi_0\right) \, &\,  \phi \,  > \, \phi_0\\
                   V_{0}+A_-\, \left(\phi-\phi_0\right),\, & \,\phi \, < \phi_0
                \end{array},
         \right.
\end{eqnarray}
where $A_-$ and $A_+$ are model parameters assumed to be greater than $0$. It can be found that if the width $\Delta \phi \approx (\phi-\phi_0)$ of the singularity is small enough then the resulting adiabatic primordial spectrum 
is non-flat around the point $k_c$ which can be expressed analytically in terms of transfer function applied on any underlying power spectrum: 

\begin{equation}
\mathcal{P}(k) \, =\, \mathcal{P}_o(k) \, \mathcal{D}^2(y,\Delta),
\label{staropk}
\end{equation}
where the transfer function is given by \cite{2012JCAP...01..008M,2014JCAP...07..014C},
\begin{eqnarray}
\mathcal{D}^2(y,\Delta) \, &=& \left[1+\frac{9\Delta^2}{2}\left(\frac{1}{y}+\frac{1}{y^3}\right)^2+\frac{3\Delta}{2}\left(4+3\Delta-\frac{3\Delta}{y^4}\right)
\frac{1}{y^2}\cos(2y) \right.\nonumber \\ 
&& \left.  +3\Delta \left(1-(1+3\Delta)\frac{1}{y^2}-\frac{3\Delta }{y^4}\right)\frac{1}{y}\sin(2y)\right].
\end{eqnarray} 
Here $y=k/k_c$ and $ \Delta = \frac{A_+-A_-}{A_+}$. In this model we have applied transfer function on simple power law model $\mathcal{P}_o(k)$. $k_c$ determines the location of the step and has no effect on the shape of 
the spectrum besides the overall normalization.  For $R=A_+/A_- < 1$, there is a sharp decrease of spectrum followed by a bump at small $k$ with large oscillations and a flat upper plateau on small scales (see also \cite{2011PhRvD..83b3526G}). 
For $R > 1$ this picture is inverted and has a step-down like feature. Here, in this model for large $k$ the modulation term becomes close to 1 and we get simple power law model.
\subsection{Model 8 : Starobinsky cut off   (SBC)}
As discussed in the previous section, Starobinsky's transfer function can be imposed on any class of primordial power spectrum. Here, in this model, we superimpose 
Starobinsky modulation on an exponential cut off spectrum (model 6) with a adjustable sharpness of the cut off:  

\begin{equation}
\mathcal{P}(k) \, =\,  \mathcal{P}_o(k)\, \left[1-e^{-({\varepsilon} \, {k/k_c})^{\alpha}}\right] \, \mathcal{D}^2(y,\Delta),
\end{equation} 
where $\mathcal{D}^2(y,\Delta)$ is the transfer function of the Starobinsky feature described in the previous section and $\varepsilon$
sets the ratio of the two cut off scales involved. This model has  both exponential (sharp) cut off and a Starobinsky model bump like feature in the power spectrum. 
Previously Sinha \& Souradeep \cite{2006PhRvD..74d3518S} have found that this model provides  best likelihood value among the wide range of models discussed here.
For simplicity, we reduce the degree of freedom of this model by fixing $\varepsilon$=1. We found that this parameterization does not affect final results.

\section{Methodology and parameter estimation}
We employ  Monte Carlo Markov Chain (MCMC) analysis to estimate the parameters of PPS models we consider for
our study and use publicly available code {\tt COSMOMC} \cite{2002PhRvD..66b3531L,cosmomc}  for this purpose.
{\tt COSMOMC} uses publicly available code {\tt CAMB} \cite{2000ApJ...538..473L,camb} for computing angular power spectra of CMB 
anisotropies following a line of sight approach which was given in \cite{1996ApJ...469..437S}. {\tt COSMOMC} uses the likelihood
code provided by the WMAP and Planck team for computing the likelihood.

WMAP 9 parameter estimation methodology is given in \cite{2013ApJS..208...19H} which is not very different 
than what was outlined in \cite{2003ApJS..148..195V}.  WMAP 9 year likelihood code does not need any extra
parameter and computes the likelihood at low and high $\ell$'s differently for the temperature and 
polarization data.  At low-$\ell$ ($l \le 32$) TT likelihood is computed from the angular power 
spectrum estimated using Gibbs sampling and at high-$\ell$ ($l > 32$) TT likelihood is calculated from the angular power 
spectrum  estimated from an optimum quadratic estimator. 
For polarization, high-$\ell$ ($l > 23$) TE, EE and BB likelihoods are computed using {\tt MASTER} and low-$\ell$ ($ l \le 23$),
TE, EE and BB likelihoods are computed in the pixel space. 

Apart from WMAP 9 year temperature and polarization data, we also consider Planck temperature data
for our analysis which is also publicly available. 
Planck likelihood code (discussed in \cite{2014A&A...571A..15P} and  downloadable from \cite{plc})  
has different modules to compute likelihood at low and high $\ell$'s. 
Planck likelihood code also computes likelihood for low-$\ell$ polarization data which 
it uses from  WMAP 9 year data, however, we do not use that.  We consider only
modules which compute TT-likelihood at low and high $\ell$. At high-$\ell$ (up to $\ell$=2500),
Planck likelihood code uses a code named {\tt CamSpec} which has 14 extra parameter 
to take care of foreground and other systematic. At low-$\ell$ ($\ell$ $\leq$ 49) Planck likelihood
code uses {\tt COMMANDER}.

Since we perform joint WMAP 9 + Planck analysis of CMB power spectrum, it is important to discuss about consistency 
between WMAP 9  and Planck data. Within in the context of the $\Lambda$CDM model, it has been found that the values of 
some cosmological parameters like $H_0$ obtained from Planck measurements are significantly different from WMAP 9 measurements.
It was shown in \cite{kovacs2013} that WMAP 9 angular power spectrum is about 2.6\% higher at very high significance level at low $\ell$'s, 
however, no significant bias was found at high $\ell$'s. Similarly, Planck team has reported the the inconsistency of the angular power spectrum at  multipoles $\ell \le 40$ \cite{2014A&A...571A..15P} and 
2\% difference of angular power spectra near the first first acoustic peak \cite{2014A&A...571A..31P}. It was also  shown by  \cite{2014PhRvD..89d3004H} that although best fit model to Planck data was consistent with WMAP 9 year data, but 
WMAP 9 best fit was found to be inconsistent with Planck data at 3$\sigma$. Recently \cite{2015ApJ...801....9L}  revisited the analysis of the WMAP 9 data and they also found  found ∼2.5\% difference 
in WMAP 9 and Planck spectra at $\ell \ge100$ at 3-5$\sigma$ level. Although, some level of the inconsistency between cosmological measurements was found arising from the
Planck $217\times217$GHz detector \cite{10.1103/PhysRevD.91.023518}, but there still remains significant tension. The tension between WMAP 9 and Planck data could be due the different systematics 
present in the WMAP 9 and Planck data or it could signal the the failure of 
$\Lambda$CDM model which could have far reaching consequences. 

We modify {\tt CAMB} and {\tt COSMOMC} so that the extra parameters of the PPS models
can be incorporated and use priors as are given in Tab.~(\ref{priorranges}). Since for
running {\tt COSMOMC} we need covariance matrices also apart from prior range 
therefore we generate covariance  matrices from a few trial runs.

\begin{table} 
\centering
\begin{tabular}{|l|c|c|}
\hline
Parameter Name& Symbol & Prior Ranges \\ 
\hline \hline
Baryon Density & $\Omega_b{h}^2$ & 0.005-0.1 \\ 
Cold Dark Matter Density & $\Omega_{c}{h}^2$ & 0.001-0.99 \\ 
Angular size of Acoustic Horizon& $\theta$ & 0.5-10.0 \\
Optical Depth & $\tau$ & 0.01-0.8  \\ 
Scalar Spectral Index & $n_s$  &  0.5-1.5 \\ 
Scalar Amplitude & log$10^{10} A_s$ & 2.7-4.0\\
Hubble Parameter at Inflation &$H_{inf}$ (Mpc$^{-1}$)& $10^{-2}$-$10^{-7}$\\ 
Running Index    & $\alpha_s$  & $-$1-1\\
Cut off Parameter  & $k_{c}$ (Mpc$^{-1}$) & 0.0-0.01 \\
Cut off Steepness Parameter & $\alpha$ & 1.0-15.0 \\
Starobinsky Parameter & $\Delta$ & 0.0-1 \\ 
\hline
\end{tabular}
\caption{Uniform prior used in parameter estimation.}
\label{priorranges}
\end{table}
The cosmological parameterization has been carried out by using 
the six basic parameters (baryon density ``$\Omega_bh^2$", cold dark matter density ``$\Omega_{c}h^2$'', Thomson scattering optical depth due 
to reionization ``$\tau$'',  angular size of horizon ``$\theta$'', spectral index ``$n_s$'' and scalar amplitude 
``$\ln10^{10}A_s$'') along with the parameters which describe the features in the PPS i.e., $k_c$, $\alpha$, $\Delta$ etc.

Apart from the standard six cosmological parameters, we keep the values of the rest of the cosmological parameters constant. 
We have fixed the sum of physical masses of standard neutrinos ``$\nu$''=0.6 eV, effective number of 
neutrinos ``$N_{eff}$''=3.046, Helium mass fraction ``$Y_{He}$''=0.24 and the width of reionization $0.5$. 
For the case of WMAP 9 year + Planck data all the nuisance parameters of {\tt CamSpec} where fixed to the standard 
values given in \cite{2014A&A...571A..15P,2014A&A...571A..16P}.   

We perform a Markov Chain Monte Carlo analysis to determine the values of the model parameters that provide the best fit to the observed
data from Planck and WMAP for CMB power spectrum. {\tt Getdist}  was used with the chains generated by {\tt COSMOMC} 
to produce 2D contours and plots of the marginal posteriors.  

\subsection{Best fit parameters}

We present the results of our analysis in terms of the best fit values and their mean values with 1-$\sigma$ errors 
(when possible) for the parameters of the PPS models which characterize the primordial power spectrum.
Since we find that the values of the rest of the cosmological parameters are within acceptable range and do
not show any interesting correlation with our model parameters of PPS, so we do not present estimates for those here.

We present the estimates of model parameters for the WMAP 9 year and WMAP 9 year + Planck data in 
Tab.~\ref{estimates} with the values of -2$\log$ likelihood (or ``$\chi^2$''). From the table we can see that
all  the models we consider give better fit to the data than the standard power law model. However, 
the improvement is marginal. We also present the ranking of the models later in this section.


\begin{table*}[h]
\noindent\resizebox{\linewidth}{!}{
\centering
\begin{tabular}{|c|c|ccc|ccc|}  \hline
\multicolumn{2}{|c|}{} & \multicolumn{3}{c|}{WMAP 9} &\multicolumn{3}{c|}{WMAP 9+Planck}             \\  \hline \hline
Model   & Parameter          &  Best Fit   & 68\% Limit &$\chi^2 =-2\log {\mathcal L}$                 & Best Fit   &  68 \% Limit  &$\chi^2 =-2\log {\mathcal L}$           \\ \hline
1 (PL)  &                    &             &                          &  7558.0160 &     &                             &15382.9400\\ \hline
2 (RN)  & $\alpha$            & -0.012        & -0.012$\pm$ 0.022      &7557.7340  &  -0.009     &  -0.009 $\pm$ 0.006  &15380.6580 \\ \hline
3 (SC)  & $10^4 k_c$         &  2.9149       &  2.3597$\pm$ 0.9809   & 7555.6080 &   3.0449   & 2.5653$\pm$0.8250       &15378.2840\\ \hline
4 (PIR)  & $10^4k_{c}$          & 0.3910      &  0.5909$\pm$0.5324    & 7557.9100 &  0.3941    &  $<$0.4296                 & 15382.1560\\ \hline
5 (PIK)  & $10^4H_{inf}$       & 2.0846      & 2.07836 $\pm$ 1.0052  & 7556.1900  & 2.1485     &  2.0934$\pm$ 0.8973  & 15380.0950\\ \hline 
6 (EC) & $10^4k_c$           & 2.9244      & 2.4876$\pm$1.1406     &7555.6700   & 2.9780    &  2.7752$\pm$ 0.9237        &15378.6420 \\ 
       & $\alpha$            & 7.6167      &  8.1354[NL]           &           & 9.22328    &  8.2764[NL] &                            \\ \hline
7 (SB) & $10^4k_c$           & 1.4724      &   $<$ 12.83404    & 7556.1760 &   14.641    &  $<$ 14.5739                     &15375.7760 \\ 
       & $\Delta$            & 0.3893      &   $<$ 0.2583   &             &   0.0558    & 0.0696$\pm$0.0667   & \\                      \hline
8 (SBC) & $10^4k_c$           & 3.1313      & $<$  0.1839    & 7555.7640   &  2.9149     & $<$  0.23354                     &15378.3460\\ 
       & $\alpha$            &12.502       & 8.022[NL]    &             &  12.6627     & 8.1238[NL]  &                           \\ 
       & $\Delta$            &0.0037       & $<$0.4509     &             &  0.055492    & $<$0.2896  & \\                         \hline

\end{tabular}
}
\caption{The best fit and mean values of the extra parameters of PPS models we consider for the WMAP 9 year
and joint WMAP 9 year and Planck data. We find good constrains on the cut off scales $k_c$, however,
our constrains on other parameters are poor. We were able to put upper limit on the Starobinsky parameter $\Delta$ 
but no limit (NL) was found for the exponential cut off parameter $\alpha$.
}
\label{estimates}
\end{table*}

\begin{figure*}[here]
\begin{center}
\includegraphics[width=9.5cm,height=9.5cm,keepaspectratio]{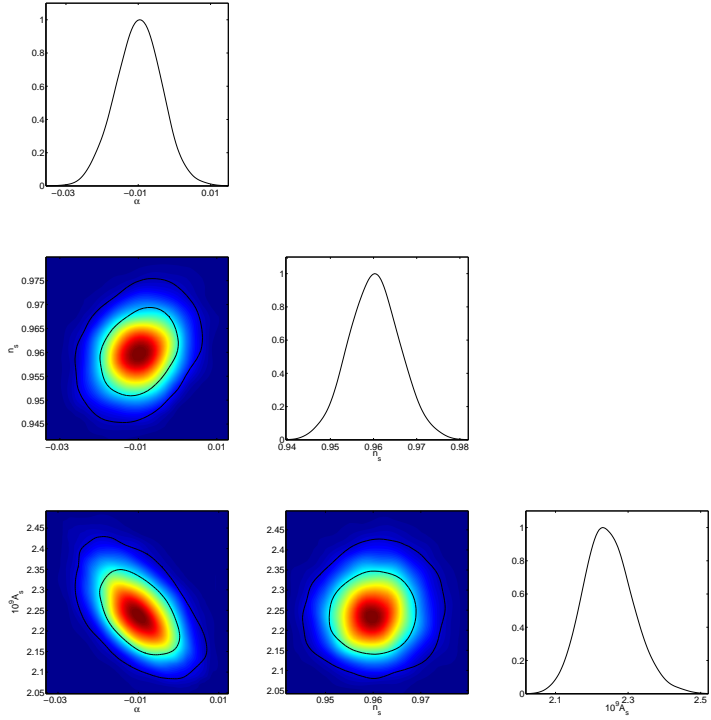}
\caption{Two dimensional joint probability distributions and one dimensional
marginal probability distribution  for the parameter $10^9A_s, n_s$ and $\alpha_s$ of
the primordial power for RN model for WMAP 9 + Planck data.} 
\label{cont_m2}
\end{center}
\end{figure*}

\begin{figure*}[here]
\begin{center}
\includegraphics[width=9.5cm,height=9.5cm,keepaspectratio]{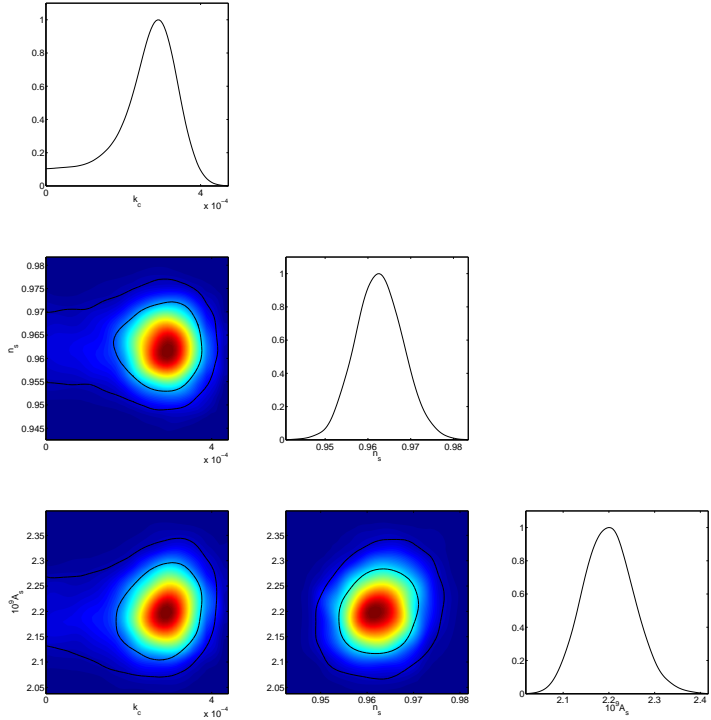}
\caption{
Same as in Fig.~(\ref{cont_m2}) for the model 3 (SC). From the figure
it is clear that the cut off scale is quite well constrained from the data.}
\label{cont_m3}
\end{center}
\end{figure*}

\begin{figure*}[here]
\centering
\includegraphics[width=9.5cm,height=9.5cm,keepaspectratio]{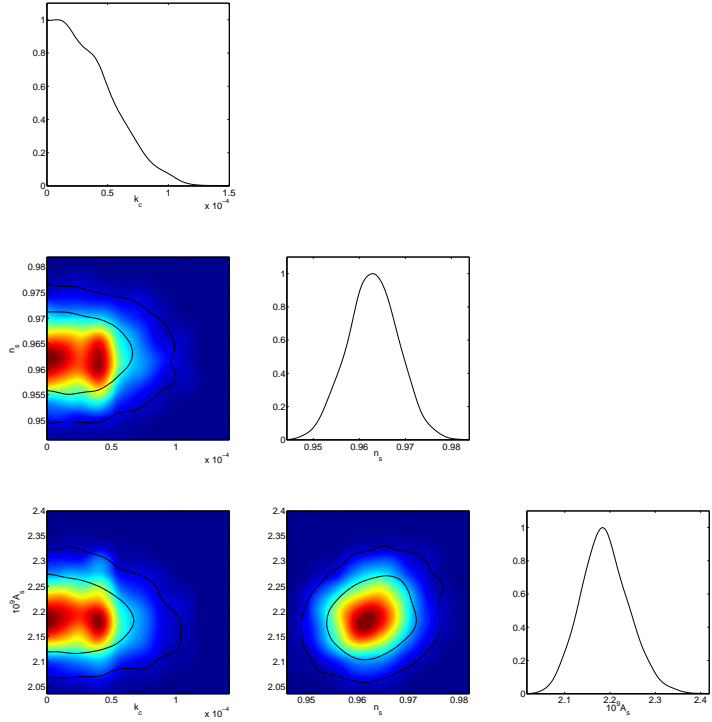}
\caption{Marginal one dimensional probability distributions and two dimensional joint probability
distribution for the model 4 (PIR) which also has just one extra parameter  for the  WMAP 9 + Planck data.
For this model the constraints on $k_c$ are not that good as we have for model 3 (SC) which
is expected since the role of $k_c$ in this case is slightly different.}
\label{cont_m4}
\end{figure*}

\begin{figure*}[here]
\centering
\includegraphics[width=9.5cm,height=9.5cm,keepaspectratio]{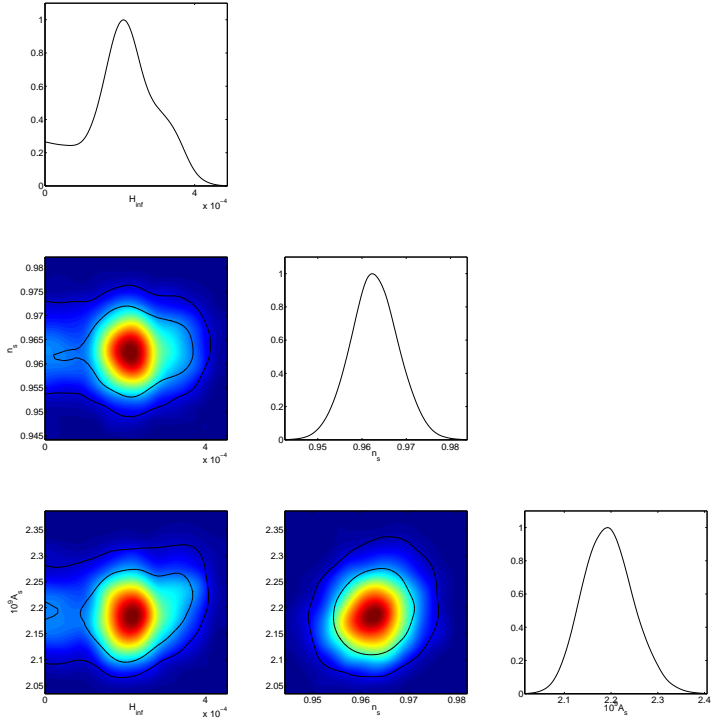}
\caption{One dimensional marginalized probability distribution and contour plots 
for the parameters of PPS for model 5 (PIK) with WMAP 9 + Planck data.}
\label{cont_m8}
\end{figure*}

\begin{figure*}[here]
\centering
\includegraphics[width=9.5cm,height=9.5cm,keepaspectratio]{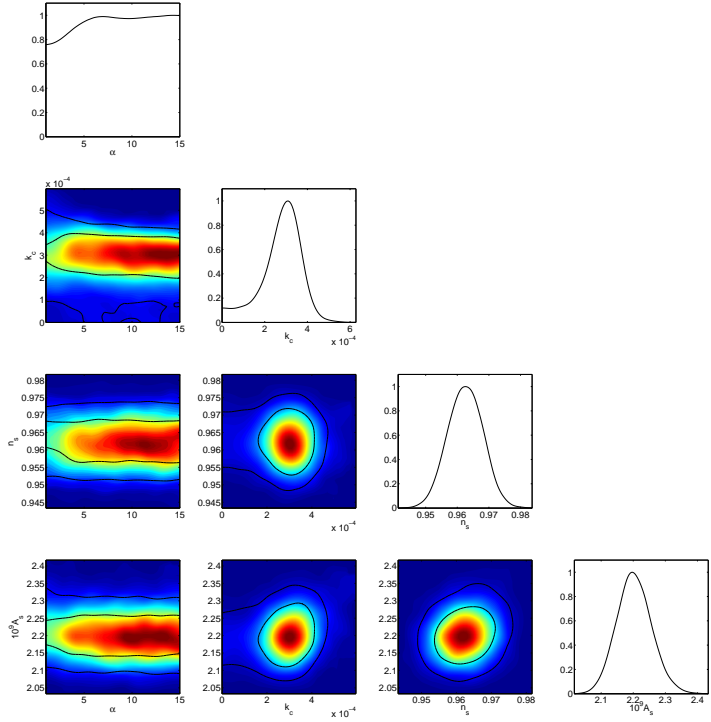}
\caption{We find that the likelihood for WMAP 9 + Planck data is
not very sensitive for parameter $\alpha$ for model 6 (EC) so we have poor constrains 
(any value of $ \alpha > 5$ value is as good as any other value). However, for this model 
we also obtain good constrains on the cut off scale $k_c$.}
\label{cont_m5}
\end{figure*}

\begin{figure*}[here]
\centering
\includegraphics[width=9.5cm,height=9.5cm,keepaspectratio]{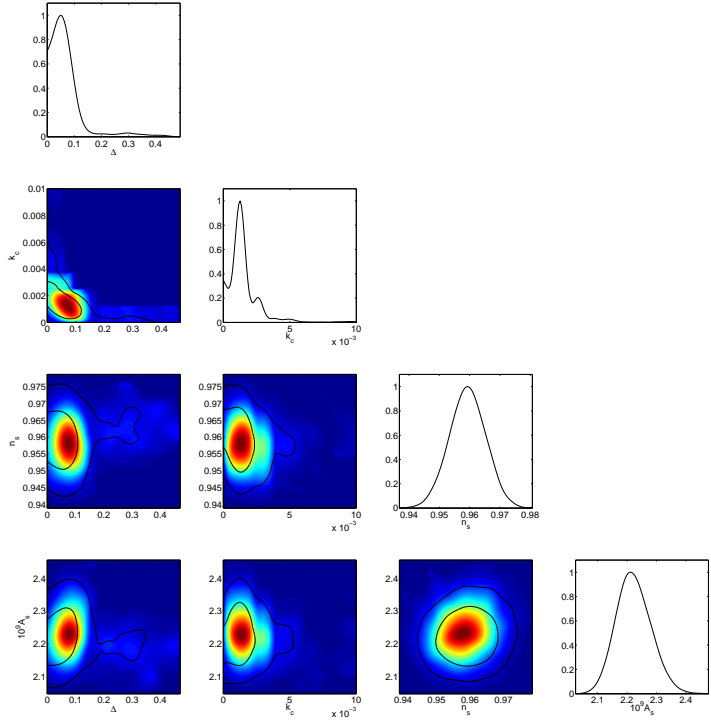}
\caption{For model 7 (SB), we find upper limits for the fitting parameters $k_c$ 
and $\Delta$ for the  WMAP 9 + Planck data.}
\label{cont_m6}
\end{figure*}

\begin{figure*}[here]
\centering
\includegraphics[width=9.5cm,height=9.5cm,keepaspectratio]{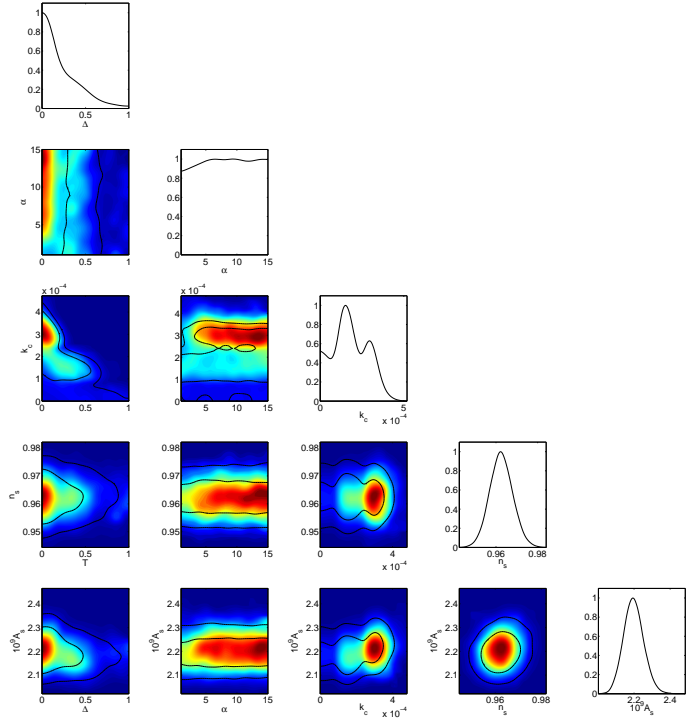}
\caption{Like model 6 (EC), we find for model 8 (SBC) poor constraints for 
$\alpha$, however, we find good constrains for the parameter $k_c$ and $\Delta$ 
for the WMAP 9 + Planck data.}
\label{cont_m7}
\end{figure*}

\begin{figure*}[here]
\centering
\includegraphics[width=9.5cm,height=9.5cm,keepaspectratio]{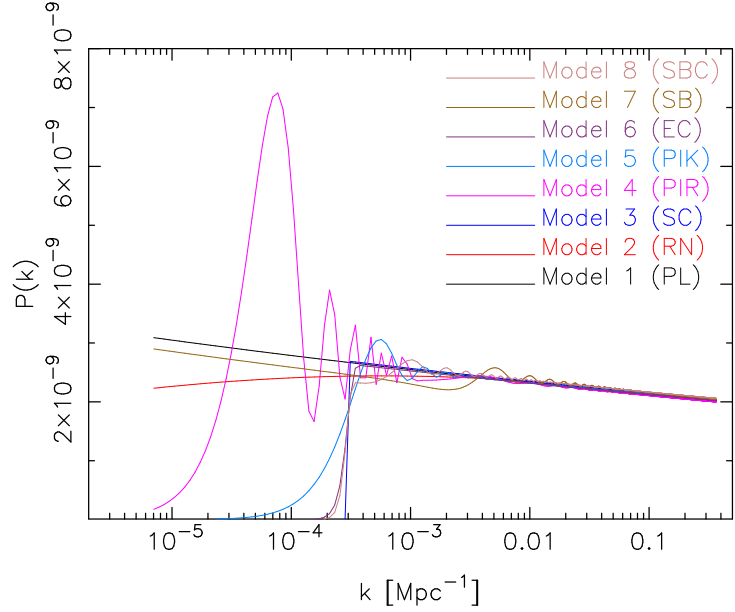}
\caption{The best fit primordial spectra for the models consider for our analysis using WMAP 9 + Planck data. 
Note that all the models we considered in this work have cut off at large scale $ k < k_c$ and matched with the standard power law model. All the models
we consider give better likelihood than the pure power law model.}
\label{bestfit_pk}
\end{figure*}

\begin{figure*}[here]
\centering
\includegraphics[width=9.5cm,height=9.5cm,keepaspectratio]{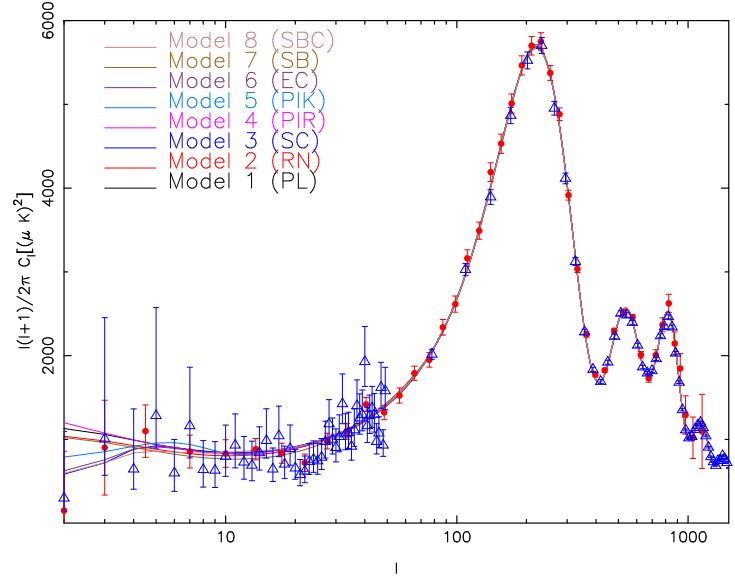}
\caption{The best fit angular power spectra $C_l^{TT}$ for the models of PPS we consider using WMAP 9 + Planck data. The observed
data points for WMAP 9 + Planck data are also shown by red dots and blue triangles  
respectively with error bars.}
\label{bestfit_cl1}
\end{figure*}

\begin{figure*}
\centering
\includegraphics[width=9.5cm,height=9.5cm,keepaspectratio]{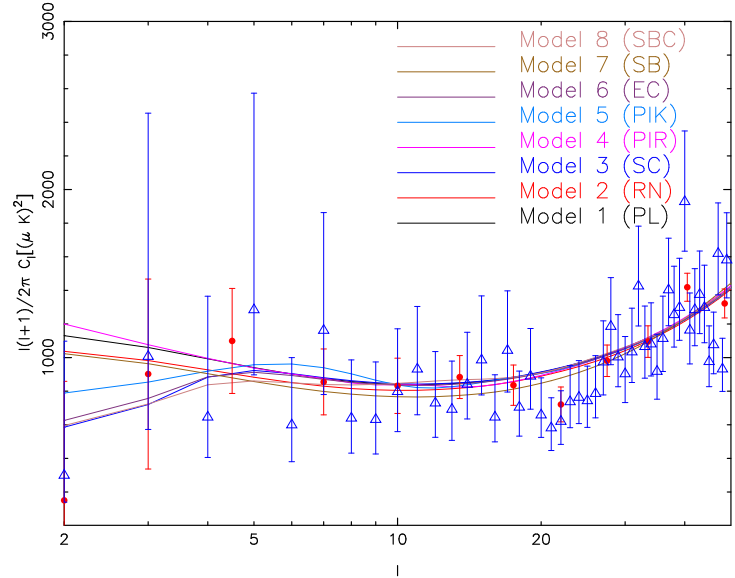}
\caption{Same as in Fig.~(\ref{bestfit_cl1}) at low-$\ell$. This figure shows that the 
model 7 (SB) gives power suppression up to a large values of $l$ so fit the Planck data
better than other models.}
\label{bestfit_cl2}
\end{figure*}

One of the features which all of our models (apart from PL) have common is a cut off at large scale we have and
characterized by a scale $k_c$. We find that cut off around $ 1.40\times 10^{-4}-3.15 \times  10^{-4} \text{Mpc}^{-1}$ for most of the  models 
discussed here, except for model (4) which predicts much smaller value of $k_c$ for both data sets and model (7) which predicts much larger values of $k_c$
 for joint WMAP 9 year + Planck data set. Fig's.~ (\ref{cont_m2}), (\ref{cont_m3}), (\ref{cont_m4}), (\ref{cont_m8}),
(\ref{cont_m5}), (\ref{cont_m6}) and (\ref{cont_m7})   show the marginal one-dimensional posterior distributions and 2D contours at 68\% and 95\% CL of the parameters describing 
PPS models discussed in the work. For models (3), (6), (7), (8), we were able to obtain good bounds on the $k_c$ for the WMAP 9  + Planck data as is clear from their
two dimensional joint probability distributions. Fig. (\ref{bestfit_pk}) shows the WMAP 9  + Planck best fit primordial spectra for the range of models discussed in this work. 
Note that all the models we consider here have cut off at large scale $k < k_c$ and matched with the standard power law model at large $k$. 
Fig's.~(\ref{bestfit_cl1}) \& (\ref{bestfit_cl2}) show the corresponding angular power spectra $C_l^{TT}$ obtained using best fit values of PPS parameters and other standard cosmological parameters.

\subsection{Model comparison}


There are many methods for model comparison and have their own advantages and disadvantages. Two of the 
most common methods are Akaike Information Criteria or $AIC$ \cite{Akaike1974_aic} and Bayesian 
Information Criteria or $BIC$ \cite{Schwarz1978_bic}. $BIC$ is considered more powerful than $AIC$
since it explicitly takes into account the number of data points when penalizing the models with 
more  parameters.  $AIC$ has been most commonly used \cite{2004MNRAS.351L..49L,2014JCAP...07..014C} 
and is  defined in the following way:

\begin{equation}
AIC = -2 \ln {\mathcal L}({\bf d}|{\bf \theta}) + 2 k, 
\label{aic_def}
\end{equation}
where ${\mathcal L}({\bf d}|{\bf \theta}) $ is likelihood, ${\bf d}$  data vector, $\theta \in {\mathbb R}^k$ is the parameter vector and $k$ is the number of parameters. 
The best model is one which have minimum value of $AIC$. $BIC$ is also defined in the similar  way:
\begin{equation}
BIC = -2 \ln {\mathcal L}({\bf d}|{\bf \theta}) +  k \ln N, 
\label{bic_def}
\end{equation}
where $N$ is the number of data points. 

The second terms in the RHS of equation~(\ref{aic_def}) and (\ref{bic_def}) work as 
regularization functions and penalize models with more number of parameters like
any other regularization function does and can be related to Maximum Entropy Method \cite{2013PhRvD..88b3522G}. $BIC$ is more conservative than $AIC$ since it puts  stringent penalty for models
with more fitting parameters  for $ \ln N >   2$  i.e., the penalty term  exceeds $ 2k $.
It is well known that $AIC$ minimizes the Kullback-Leibler divergence between 
the estimate from a candidate model and  the true distribution 
and $BIC$ selects  a model that maximizes the posterior probability distribution.
For $AIC$, $\Delta AIC \equiv AIC_{i}-AIC_{min}$ represents preference 
of model $i$ over the the best fit model. Models with $\Delta AIC \leq 2$ have substantial support, models with $4 < \Delta AIC <7$ have considerably less support and 
those with $\Delta AIC > 10$ have essentially no support compared to best fit model \cite{Kenneth2002}. $\Delta BIC \equiv BIC_{i}-BIC_{min}$ represents the preference of best fit model ( i.e model with 
 minimum value of $BIC$ ($BIC_{min}$)) 
over model $i$. $\Delta BIC$ values of $\Delta BIC\le2 $, $2< \Delta BIC\le 6$, $6<\Delta BIC\le 10$ and $\Delta BIC> 10$ 
represent weak, positive, strong and very strong support for best fit model respectively \cite{Kass1995}. However, in order to calculate $BIC$, it is not straightforward 
to find out the number of independent data points ($N$)  for WMAP 9 year + Planck data as it is a correlated data set.
But it is clear that as the number of data points is quite large, $BIC$ becomes highly sensitive towards the number of extra parameters.
For a sensible estimates of $N$, we can assume 1168 (WMAP 9 Master TT $\ell$'s) +  777 (MWAP 9 Master TE $\ell$'s)
+  1170 (WMAP 9 TT/TE/EE/BB low-$\ell$ chi2 pixels) + 2499 (Planck CamSpec+commander $\ell$'s) = 5614 
independent data points for WMAP 9 + Planck data set. Therefore, we get for WMAP 9 data $\ln (N) = 8.04$ and  $\ln (N) = 8.63$ for WMAP 9 + Planck data set.

\begin{table*}[h]
\centering
\begin{tabular}{|c|c|c|c|c|}  \hline
\multicolumn{1}{|c|}{} & \multicolumn{2}{c|}{WMAP 9} &\multicolumn{2}{c|}{WMAP 9+Planck}             \\  \hline \hline
Model   &   $ \Delta AIC$     & $ \Delta BIC$  & $\Delta AIC$   &  $\Delta BIC$     \\ \hline
1 (PL)  &  0.408      & -5.632 &  3.164      &  -10.036 \\ \hline
2 (RN)  &  2.126      & 2.126  &   2.882     &  -3.718 \\ \hline
3 (SC)  &  0.000      & 0.000  &   0.508     &  -6.092   \\ \hline
4 (PIR)  &  2.302      & 2.302  &   4.380     &   -2.22    \\ \hline
5 (PIK)  &  0.582      & 0.582  &   4.319     &  -4.281        \\ \hline 
6 (EC)  &  2.062      & 8.102  &   2.866     &   2.866         \\  \hline 
7 (SB)  &  2.56       & 8.608  &   0.000     &   0.000        \\  \hline 
8 (SBC) &  4.156      & 16.236 &   4.570     &   11.17          \\  \hline 
\end{tabular}
\caption{This tables shows $\Delta AIC$ and $ \Delta BIC$ for the different models for WMAP 9 and WMAP 9+Planck data. For WMAP 9 data 
we find that the sharp cut (SC)  model  (model 3) gives lowest $AIC$, however, for WMAP 9 year + Planck,
Starobinsky model (SB)  model (model 7) gives the lowest $AIC$. The reason behind model 7 being preferred 
by WMAP 9 year + Planck data is that it suppresses power at higher angular scales $\ell \leq30$. However, $BIC$ always prefers power law model over the 
cut off models because of least number of parameters.}
\label{aic_bic}
\end{table*}

We present $\Delta AIC$ and $\Delta BIC$ values for the models we consider
in Tab.~(\ref{aic_bic}). 
We find that for the WMAP 9 year data, sharp cut off model (SC) gives the best (minimum) value of $AIC$ but is as good as power law model, however, for WMAP 9 year + Planck data Starobinsky model (SB) has the 
lowest value of $AIC$ which we believe is this due to the fact the this model gives suppression of power up a higher values of $l$ as required by the Planck data. Using $BIC$ rule for model selection, one 
finds that power law model is always favored over the cut off models. Moreover, among cut off models, model with lesser number of parameters is always favored.
It is also worth mentioning that model (8) has a larger value of $\chi^2$ despite having more parameters and therefore is disfavored by the current data.

\section{Discussion and conclusions}

The observed CMB power spectrum is in striking agreement with the standard  $\Lambda$CDM model with  almost 
scale-invariant adiabatic fluctuations produced during the inflationary epoch. 
 However in such studies, some anomalies have been observed such as low CMB power on large angular scales. 
It has been observed that inflationary epoch cannot be well described by simple form of scalar power spectrum 
based  on the smooth slow roll approximation and the presence of the cut off in the primordial power spectrum 
is  essential for extension of this simplistic picture in order to explain low CMB anomaly. 
In this work, we have explored different parameterizations of inflationary driven primordial 
spectra  which have cut off at large angular scales so as to describe low CMB anomaly.

We have analyzed the complete CMB data sets of WMAP 9 year and Planck. We perform a Markov Chain Monte Carlo 
analysis to determine parameters that provide the best  fit to the data for the CMB angular power spectrum. 
We find that primordial power spectrum with cut off, in general, leads to improvement 
of the likelihood and marginal preference for a non vanishing cut off scale of $k_c$ in some cases.  
Due to the large variance in the CMB  temperature at low multipoles, we could only place weak constraints on  some parameters of our model like $\alpha$, however, 
our constrains on the cut off scale are fairly good.

In order to quantify the significance of the fits we have used Akaike information criterion and Bayesian information criterion.
We find that for the WMAP 9 data, among various models discussed here,  model (3) which has sharp
cut off provides  best likelihood value, but is as good as power law model as per $AIC$.  For the WMAP 9 year + Planck data set, we find only Starobinsky's model (7) 
is able to explain suppression up to multipoles $\ell \leq 30$  as is indicated by the much large value of the $k_c=14.64\times 10^{-4}$ Mpc$^{-1}$ and is slightly preferred over other models as per $AIC$.
Although, Starobinsky’s model improves the fit in the Planck CMB power spectrum in the region $\ell \leq 30$, the produced suppression is still not enough and there is some scope of
improvement in the fit. However, using $BIC$, one finds that power law model is always preferred over all cut off models.  

It is also important to note that the power suppression in the CMB anisotropy is currently a subject of intense debate.
The CMB suppression could be caused by other mechanisms and therefore, it is important to improve and develop current 
techniques which allow us better understanding of CMB suppression.
There are also models such as \cite{2001PhRvD..64j3502E,2010JCAP...06..009F,2012PhRvD..85b3531A} which have oscillations 
superimposed on primordial power spectrum which also provide better fit to CMB power spectrum, but 
in these models the cut off is not evident as is preferred by current data and oscillations cover over whole range of $k$ 
(unlike the models discussed here, where oscillations die out after the cut off).
Finally, we conclude that the present motivation for the low CMB power at small $\ell$ with an infrared cut off is very high and 
there is significant scope in improving the estimates of power suppression 
on the basis of modeling of the primordial power spectrum with the upcoming and future data. 
\section*{Acknowledgments}

This work was supported by DST Project Grant No. SR/S2/HEP-29/2012. JP would like to thank the Science and Engineering Research Board (SERB)
of the Govt. of India for  financial support  via a Start-Up Research Grant (Young Scientists) SR/FTP/PS-102/2012. AI and MM  would like to thank IUCAA for its hospitality during their stay. 
We acknowledge the use of IUCAA's High Performance Computing facility for carrying out this work. We are thankful to WMAP and Planck teams for making data and other important softwares publicly available.

\bibliographystyle{JHEP}
\bibliography{pps}

\end{document}